\documentclass[linenumbers,trackchanges]{aastex701}

\usepackage{booktabs}
\usepackage{xcolor}
\usepackage{amsmath,amssymb,CJK}
\usepackage{gensymb}
\usepackage{soul}
\usepackage{ulem}
\usepackage{threeparttable}

\newcommand{\zx}[1]{\textcolor{black}{#1}}
\newcommand{\sho}[1]{\textcolor{black}{#1}}
\newcommand{\jn}[1]{\textcolor{black}{#1}}
\newcommand{\db}[1]{\textcolor{black}{#1}}
\begin{document}
\nolinenumbers
\setlength{\parindent}{0.25in}

\title{Water Production Rates of the Interstellar Object 3I/ATLAS}

\author[0000-0003-2399-5613]{Zexi Xing} 
\affiliation{Physics Department, Edmund C. Leach Science Center, Auburn University, Auburn, AL 36849, USA}
\email{zzx0030@auburn.edu}

\author[0000-0001-6422-1038]{Shawn Oset} 
\affiliation{Physics Department, Edmund C. Leach Science Center, Auburn University, Auburn, AL 36849, USA}
\email{szo0032@auburn.edu}

\author[0000-0003-2152-6987]{John Noonan}
\affiliation{Physics Department, Edmund C. Leach Science Center, Auburn University, Auburn, AL 36849, USA}
\email{noonan@auburn.edu}

\author[0000-0002-2668-7248]{Dennis Bodewits}
\affiliation{Physics Department, Edmund C. Leach Science Center, Auburn University, Auburn, AL 36849, USA}
\email{dennis@auburn.edu}

\begin{abstract}
We report the detection of water activity in the third confirmed interstellar object, 3I/ATLAS, based on ultraviolet imaging with the \emph{Neil Gehrels Swift Observatory}’s Ultraviolet/Optical Telescope (UVOT).
\zx{Assuming a reddening of 29\% between 3325.7~\AA\ and 5437.8~\AA, \sho{measurements} on 2025 July 31–August 1 yielded a first, marginal detection of OH (A$^2\Sigma$ -- X$^2\Pi$) emission near 3085~\AA, corresponding to a water production rate of $(0.74 \pm 0.50) \times 10^{27}$ molecules\,s$^{-1}$.
The subsequent visit on 2025 August 18–20 revealed a clear OH detection, implying a higher water production rate of $(1.36 \pm 0.35) \times 10^{27}$ molecules s$^{-1}$ (40 kg~$s^{-1}$) at a heliocentric distance of 2.90~au.}
This places 3I/ATLAS among the few comets with confirmed OH emission beyond 2.5~au, where water ice sublimation from the nucleus is typically inefficient. The inferred production rate at 2.9 au implies an active area of at least \zx{7.8~km$^2$}, assuming equilibrium sublimation. This requires that over 8\% of the surface is active, which is larger than activity levels observed in most solar system comets. Contemporaneous near-infrared spectroscopy indicated the presence of icy grains in the coma, which may serve as an extended source of water vapor. 
\end{abstract}

---

\section{Introduction} 
The discovery of the third interstellar object, 3I/ATLAS, on 1 July 2025 initiated a broad characterization campaign across the globe \citep{MPC2025K25N12,Seligman2025}. Following the lessons learned from the prior interstellar objects 1I/`Oumuamua and 2I/Borisov, observing campaigns were initiated to rapidly capture its initial brightness, morphology, lightcurves, color, and optical and near-infrared spectrum \citep{Team2019,Fitzsimmons2023}. Given the apparent brightness and early extension of the coma the production of gas was assumed and searched for \citep{Opitom2025} but not found. Characterizing the early activity of interstellar objects is essential for understanding their chemical and physical evolution during solar approach, as it possibly represents the first time the object has been significantly heated in their very long dynamic lifetimes \citep{hopkins2025different,taylor2025kinematic}. 

Inbound activity for our own Oort Cloud solar system comets is often more predictable. Hypervolatiles like CO and CO$_2$ drive activity beyond the water ice line \citep{Ootsubo2012,Pinto2022}, which can entrain small icy grains into the coma that sublimate to produce H$_2$O \citep{AH2011}. As comets approach the Sun, water increasingly dominates the observed gas composition. This is not necessarily true for all solar system comets (see C/1908 R1 (Morehouse), C/2016 R2 (PANSTARRS), and C/2009 P1 (Garradd); \citep{McKay2019, Biver2018, Bod14}, nor was it the case for interstellar comet 2I/Borisov, which transitioned from an equally mixed H$_2$O/CO coma pre-perihelion to a CO-dominated coma post-perihelion \citep{Bodewits2020, Xing2020, Cordiner2020}. The change in composition was only captured thanks to the fortunate timing of UV observations by HST and the \emph{Neil Gehrels-Swift Observatory} that monitored CO and H$_2$O production. A similar monitoring campaign was planned for the next interstellar object to establish a chemical baseline for comparison.

The Ultraviolet and Optical Telescope (UVOT) onboard the \emph{Neil Gehrels-Swift Observatory} has proven to be an excellent instrument for monitoring the OH (A$^2\Sigma$ -- X$^2\Pi$) band at 3085~\AA\ \citep{Mason2007, Carter2012,Bod14, Xing2020, bodewits2023}.  Despite its modest aperture, UVOT's sensitivity is enhanced by its position above Earth's atmosphere, free from telluric absorption and sky background, enabling deep integrations and effective stacking. Because it is flown above the atmospheric extinction, UVOT has the equivalent sensitivity of a 4-m telescope on the ground\footnote{\url{https://swift.gsfc.nasa.gov/proposals/tech_appd/swiftta_v12.pdf}}. UVOT was used to derive the H$_2$O production rate of 2I/Borisov during its flyby of the solar system in 2019/2020, and it was the dual measurements of UVOT and HST COS that showed the duality of Borisov's chemical composition around perihelion \citep{Bodewits2020, Xing2020}.

Here, we present our initial UVOT observations of 3I/ATLAS taken on July 31 and August 19, 2025  that show the onset of water activity in the third interstellar object. In the following sections we describe the observations and analysis, place our results in context with the current 3I/ATLAS literature, and compare and contrast with both the previous interstellar objects and Oort Cloud comet populations.

\begin{figure*}[ht!]
\includegraphics[width=\textwidth]{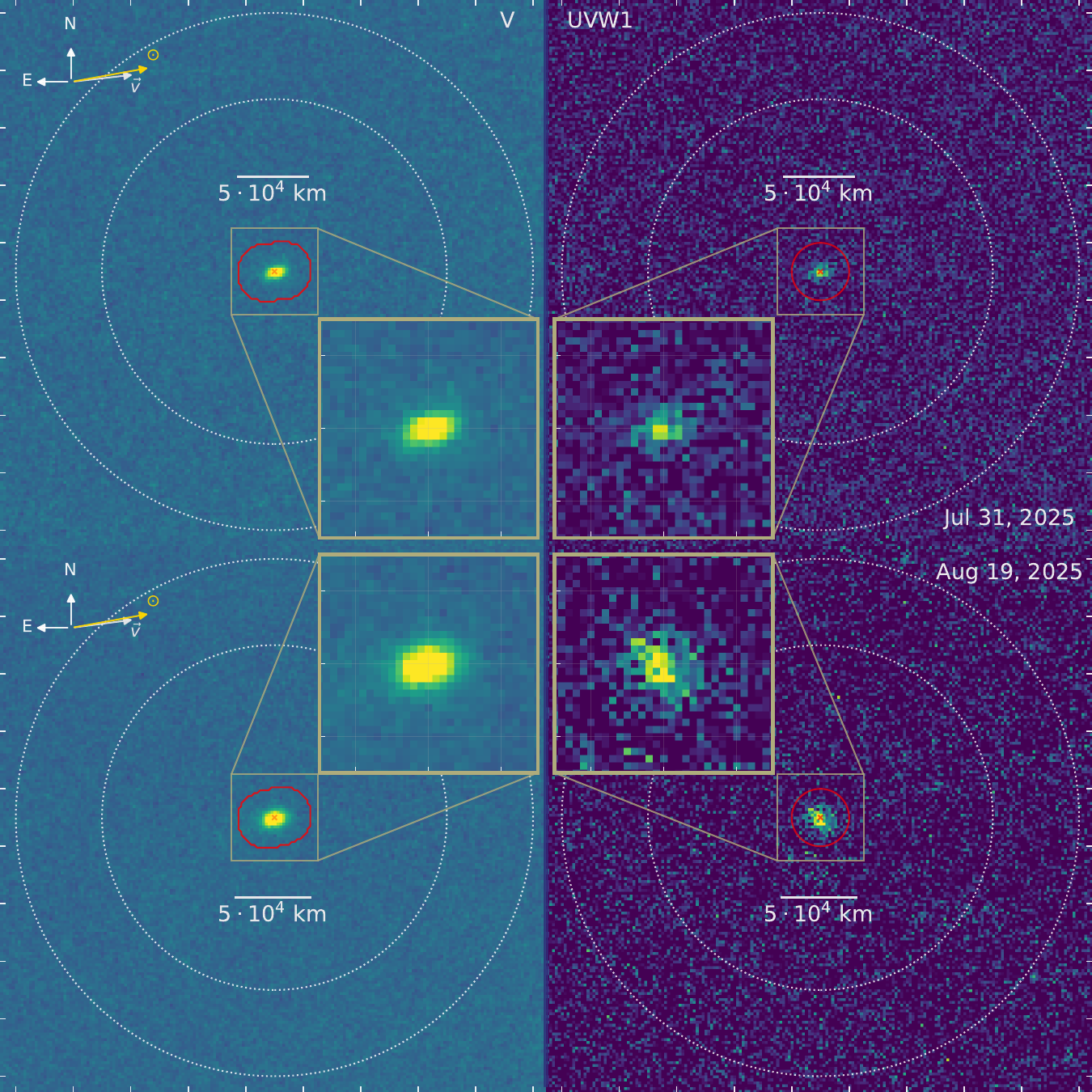}
\caption{Stacked images of interstellar comet 3I/ATLAS acquired with UVOT, 2 visits.
The first were obtained on July 31 and Aug 1, 2025 (visit 1, top half of figure), the second on Aug 19, 2025 (visit 2, bottom half of figure).
Images acquired with the V filter are shown on the left, with UVW1 images shown on the right.
Source apertures used are shown in red, with the background sky annulus in dotted white.
The UVW1 aperture is $\approx20,000$ km (10~arcsec) for Jul 31, and $\approx19,000$ (10~arcsec) km for Aug 19.
Background annulus spans 122,000 to 184,000 km for visit 1, and 114,000 to 171,000 km for visit 2.
Ticks along the edges mark 20~arcsec, with 2047 and 1908 km per arcsec for each visit respectively.
Color scales are individually optimized for viewing.
Inset grids are 10x10~arcsec, centered on the nucleus.}
\label{fig:aperture_fig}
\end{figure*}

\section{Observation and Analysis} \label{sec:style}
We observed 3I/ATLAS with the Ultraviolet/Optical Telescope \citep[UVOT;][]{Roming2000} on board the \emph{Neil Gehrels-Swift Observatory} \citep{Gehrels2004}. UVOT is a 30-cm Ritchey–Chr\'etien telescope with a 17 arcmin$\times$17 arcmin field of view and a plate scale of 0.502~arcsec/pixel, although most images are taken in `image mode' with 2$\times$2 on-board binning, resulting in an effective plate scale of 1.004~arcsec/pixel. It is equipped with 11 broadband filters and two grisms, covering wavelengths from 1600 to 8000~\AA.

\zx{We report here on observations conducted on two visits: 2025 July 31 – August 1, when the comet was at a heliocentric distance ($r_\mathrm{h}$) of 3.51~au and a geocentric distance ($\Delta$) of 2.82~au, and 2025 August 18–20, when it was at $r_\mathrm{h} = 2.90$~au and $\Delta = 2.63$~au.} (Table~\ref{tab:observation_and_results}). We used UVOT's UVW1 filter (central wavelength $\lambda_c$ = 2600~\AA, FWHM = 693~\AA), which includes most of the OH (A$^2\Sigma$ - X$^2\Pi$) emission band centered near 3085~\AA. To estimate and subtract the continuum contribution within the UVW1 bandpass, we also acquired images using the V-band filter ($\lambda_c$ = 5468~\AA, FWHM = 769~\AA).

\emph{Swift} does not track moving objects, so comets are typically observed with brief 200-s exposures. During our observations, 3I had a relatively high apparent motion of \zx{1.74 to 1.85 arcsec/min}, corresponding to a motion of about 6 arcsec in a typical 200-s exposure and thus exceeding UVOT's point spread function (PSF; 2.18~arcsec FWHM in V and 2.37~arcsec FWHM in UVW1 according to \citet{Breeveld_2010}). To avoid blurring and further dilution of the surface brightness, we acquired the UVW1 images in `event-mode',  where the arrival time of every photon is time tagged \citep{Poole2008}. Because of telemetry load constraints on \emph{Swift} and the higher sky brightness in the V-band, we were unable to use event mode for the V-band observations and instead obtained short-exposure images in `image' mode.

\begin{table*}[ht!]
\centering
\caption{Observations and Results Summary}
\label{tab:observation_and_results}
\setlength{\tabcolsep}{15pt}
\begin{tabular}{l cc}
\toprule
\toprule
\textbf{Visit} & \textbf{07-31} & \textbf{08-19} \\
\midrule
\multicolumn{3}{l}{\textit{Observational log}} \\
\midrule
Mid Time (UTC) & 2025-07-31 19:38:50 & 2025-08-19 07:52:11 \\
$T - T_p$ (days) & $-89.5$ & $-71.2$ \\
$r_\mathrm{h}$ (AU) & 3.51 & 2.90 \\
$\dot{r}_h$ (km/s) & $-56.29$ & $-54.33$ \\
$\Delta$ (AU) & 2.82 & 2.63 \\
S--T--O angle (deg) & 13.65 & 20.33 \\
UVW1 image number$^a$ & 16 (15) & 14 (14) \\
UVW1 exposure time (s)$^a$ & 9395 (8355) & 8206 (8206) \\
V image number$^a$ & 16 (13) & 14 (13) \\
V exposure time (s)$^a$ & 3100 (2522) & 2708 (2517) \\
\midrule
\multicolumn{3}{l}{\textit{Photometric Results$^b$}} \\
\midrule
UVW1 net count rate (cnts/s) & $0.11 \pm 0.02$ & $0.27 \pm 0.02$ \\
V net count rate (cnts/s) & $1.74 \pm 0.20$ & $3.75 \pm 0.20$ \\
V flux (erg/s/cm$^2$) & $ (9.01 \pm 1.02) \times 10^{-15}$ & $(19.36 \pm 1.03) \times 10^{-15}$ \\
V Magnitude & $17.10 \pm 0.13$ & $16.27 \pm 0.06$ \\
V A(0)f$\rho$ (cm) & 202 & 337 \\
\midrule
\multicolumn{3}{l}{\textit{OH Results$^c$}} \\
\midrule
Number of OH (molecules) & $(2.19 \pm 1.47) \times 10^{30}$ & $(4.21 \pm 1.07) \times 10^{30}$ \\
Q(H$_2$O) (molec/s) & $(0.74 \pm 0.50) \times 10^{27}$ & $(1.36 \pm 0.35) \times 10^{27}$ \\
Min. effective area (km$^2$) & 10.6 & 7.8 \\
Min. active fraction (\%) & 11 & 8 \\
Water loss mass (kg) & 22 & 40 \\
\bottomrule
\end{tabular}
\begin{tablenotes}
\small
\item{$a$}. Numbers in parentheses denote the equivalent number of images and exposure time used for analysis after excluding images contaminated by stars.
\item{$b$}. All photometric values in this table are measured within a 10-arcsec aperture; the V-band measurements used an elongated aperture of the same effective size to account for motion blur.
\item{$c$}. All OH-related results in this table are derived assuming a reddening of 29\%.
\end{tablenotes}
\end{table*}

To mitigate motion blur, we divided photons in each event-mode UVW1 image into 30-second time slices, limiting motion within each slice to less than 1~arcsec. We then aligned all slices by the target's nucleus position (from JPL/Horizons) and summed them to produce motion-corrected UVW1 images. These UVW1 images have a pixel scale of 0.502 arcsec~pixel$^{-1}$, and we binned them by 2×2 to match the 1.004 arcsec pixel$^{-1}$ scale of V-band images obtained in the image mode for subsequent image subtraction. For V-band images, such motion correction was not possible without event-mode images. 

Next, we discarded all images in which the target nucleus was contaminated by background stars. \zx{Each remaining image was divided into 50×50-pixel subregions, and stars located beyond 30~arcsec from the nucleus were identified in each subregion using a 3-$\sigma$ threshold above the local background, requiring $\ge 3$ connected pixels. Each region was then expanded by 2~pixels to cover the PSF. Within 30~arcsec of the nucleus, contaminated areas were flagged manually. All identified regions were filled with the surrounding median.}

We aligned the remaining images on the nucleus' position and stacked them separately for each filter to improve the SNR ratio. In the stacked images, we noticed that the target positions deviated from JPL/Horizons predictions. This offset, also present for stars in raw images, likely stems from a small error in the astrometric solution produced by the UVOT pipeline. The UVW1 stacked images showed a larger offset (approximately 5~arcsec) than the V images (approximately 1~arcsec), suggesting greater uncertainty in event mode. We applied a 2D Gaussian  fit to the surface to locate the nucleus position for further analyses.

\zx{Fig.~\ref{fig:aperture_fig} shows the resulting stacked UVW1 and V-band images after applying offset corrections.} All images show a clear detection of 3I/ATLAS. In the V-band images, the comet appears elongated along the direction of motion, as expected. The UVW1 images, free from smearing, show a spatially extended source. 

\zx{We acquired UVW1 photometry using a circular aperture with a radius of 10~arcsec, corresponding to 20,430~km and 19,084~km for the July 31 and August 19 visits, respectively. An elongated 10-arcsec aperture was adopted for the V-band images to account for the motion-induced elongation of the comet. To estimate the background level and its uncertainty, we randomly selected 2,000 points within an annulus 60–90~arcsec from the nucleus in each image. Around each point, we defined a region matching the size and shape of the source aperture and measured its total brightness within the region. The mean of these measurements was taken as the total background within the source aperture, and their standard deviation as the 1-$\sigma$ uncertainty of the background measurement. \ref{tab:observation_and_results}}

\begin{figure*}[ht!]
\centering
\includegraphics[width=\textwidth]{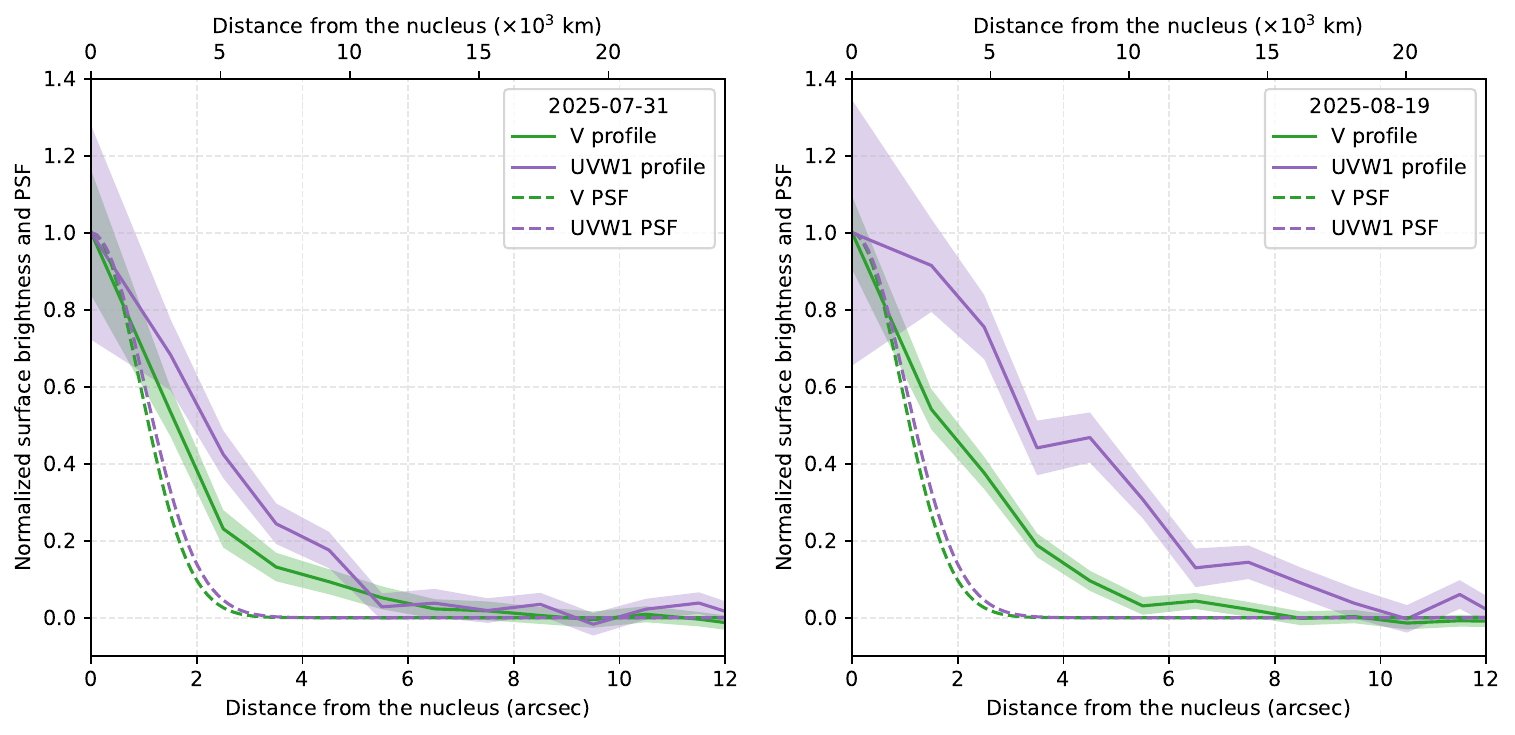}
\caption{\zx{Radial surface brightness profiles of 3I/Atlas in the V and UVW1 images for two observation visits. The left panel corresponds to the July 31 visit, and the right panel to August 19. In each panel, solid lines represent the measured profiles, while shaded regions indicate 1-$\sigma$ stochastic uncertainty. Dashed lines represent the instrumental PSF for each filter. All profiles are normalized to the brightness at the nucleus. The profiles for the V filter are represented in green, while the UVW1 profiles are depicted in purple.}}
\label{fig:profile}
\end{figure*}

\zx{We also derived background-subtracted images, from which we determined the surface brightness profiles of the coma. For the UVW1 images, the profiles were measured using a series of concentric annuli with a width of 1~arcsec, taking the mean pixel brightness within each annulus. To mitigate motion blur in the V-band images, we measured the profiles from an unaffected semicircular region confined to the sunward end. The resulting profiles, shown in Fig.~\ref{fig:profile}, are normalized to the brightness at the nucleus of each profile. The comet is clearly extended beyond the PSF in each image. In addition, in both epochs, the UVW1 profile is broader than that of the V band, confirming the presence of an additional gas contribution in the UVW1 filter.} 

We converted the V-band net count rate to band-integrated flux and magnitude using \emph{UVOT's} effective area (corrected for its ongoing sensitivity decline; \citet{Breeveld_2011_sensitivity}) and a target spectrum model, for which we adopted a linear reddening with the default solar spectrum included in the sbpy package \citep{Mommert2019}. We used a radius of 15~cm to derive the UVOT collecting area. The calculated flux, together with the target spectrum model and Vega spectrum \citep{Bohlin2014_CALSPEC}, was used to derive the apparent Vega magnitude in the V band of UVOT. The derived V-band fluxes and magnitudes are nearly insensitive to different reddening assumptions, with result variations less than 0.4\%. Therefore, the uncertainty introduced by the reddening assumption is negligible for the photometry.

The net count rate of the V image also enables an estimate of the reflected continuum contribution to the UVW1 flux, allowing us to isolate the OH (A$^2\Sigma$ -- X$^2\Pi$) emission:
\begin{equation}
    CR_\mathrm{OH} = CR_\mathrm{UVW1} - \beta(S) \cdot CR_\mathrm{V}
\end{equation}
where $CR_\mathrm{OH}$ is the count rate attributable to OH, $CR_\mathrm{UVW1}$ and $CR_\mathrm{V}$ are the count rates measured in the UVW1 and V images, respectively, as described above, and $\beta(S)$ is the solar continuum count rate ratio between the filters. The derived OH count rate depends heavily on the ratio factor $\beta$, which is a function of the reddening $S$ per 1000~\AA\ between the photon-weighted effective wavelengths of V and UVW1 filters. These effective wavelengths, calculated using the sensitivity-corrected effective area and a solar spectrum, are 5437.8~\AA\ for V and 3325.7~\AA\ for UVW1. To determine the count rate ratio $\beta(S)$, we convolved a solar spectrum linearly reddened by $S$ with the respective effective areas. 

The OH count rates are converted to the emitted flux using a hydroxyl spectral model \citep{Bod19}, then to luminosity using the geocentric distance. We subsequently derived the total number of OH molecules within the 10-arcsec radius aperture using fluorescence efficiencies \citep{Sch88} appropriate for the specific heliocentric velocities, with a $r_\mathrm{h}^{-2}$ scaling applied for the heliocentric distance.

To derive water production rates from the total number of OH molecules in the aperture, we applied the vectorial model with a random OH ejection kick of 1.05~km~s$^{-1}$ by taking photodissociation of H$_2$O into account \citep{Fes81}.
We adopted a H$_2$O outflow velocity of $0.85\times r_\mathrm{h}^{-0.5}$~km s$^{-1}$, photodissociation lifetimes of 86,000~s for H$_2$O and 129,000~s for OH, and a branching ratio of 0.93 for water dissociation into OH \citep{Com04}. We compared observed and model-predicted OH molecular numbers within the same aperture. The water production rates were then obtained by scaling the modeled water production rate by this ratio.

\section{Results} \label{sec:floats}

The measured V magnitude was $17.10 \pm 0.13$ mag (V flux = $(9.01 \pm 1.02) \times 10^{-15}$ erg s$^{-1}$ cm$^{-2}$) on July 31, and $16.27 \pm 0.06$ mag (V flux = $(19.36 \pm 1.03) \times 10^{-15}$ erg s$^{-1}$ cm$^{-2}$) on August 19. 
These UVOT V magnitudes correspond to Johnson V magnitudes of $17.07 \pm 0.13$ and $16.24 \pm 0.06$, respectively, yielding $Af\rho(0)$ values of 202 cm and 337 cm after correcting for phase angle \citep{AHearn1984,Schleicher2011}, broadly consistent with other reports of $Af\rho(0)$ measured between July 31 - Aug. 19 \citep{Manzano2025, SantanaRos2025}.

\begin{figure*}[ht!]
\centering
\includegraphics[width=\textwidth]{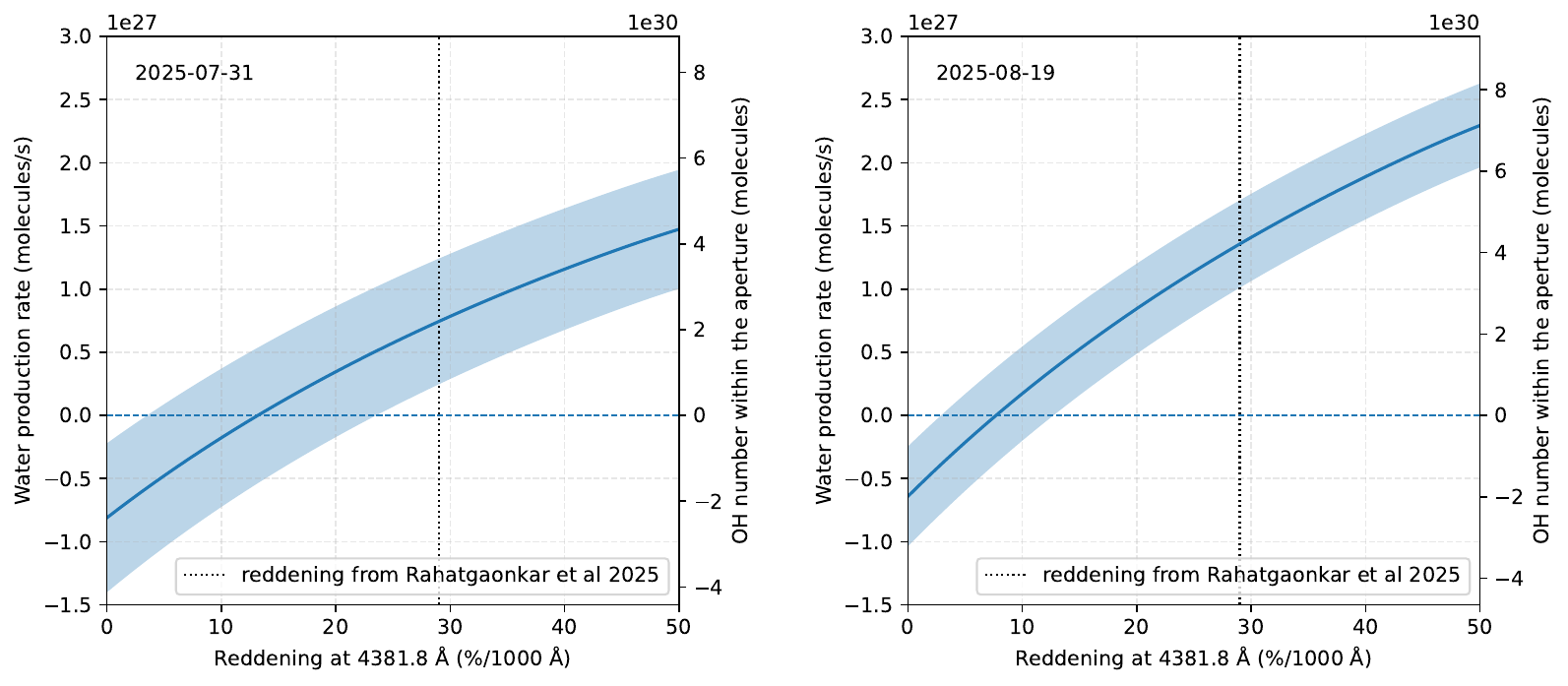}
\caption{\zx{Dependence of number of OH molecules within the 10-arcsec aperture and water production rate on reddening. The left panel presents results from the July 31 visit, while the right panel shows results from the August 19 visit. In each panel, the blue line shows the number of OH molecules and corresponding water production rates as a function of assumed reddening between V and UVW1 effective wavelengths. Shaded regions indicate 1-$\sigma$ stochastic uncertainty. The black dotted line indicates the reddening measured at the same wavelength range using reflectance spectra from \citet{rahatgaonkar2025} (private communication). The blue dashed line denotes zero production rate.}}
\label{fig:qwater}
\end{figure*}

\zx{Both the number of OH molecules within the aperture and the derived water production rate are highly sensitive to the assumed reddening between the effective wavelengths of the V and UVW1 filters (5437.8 Å and 3325.7 Å, respectively). This dependence is shown in Fig.~\ref{fig:qwater}, where 4381.8 Å in the x-axis label marks the midpoint between the two filter wavelengths. The error bars indicate 1-$\sigma$ stochastic uncertainties. During the July 31 observation, a marginal water signal (SNR$=1$) appears only when reddening assumptions exceed 24\%, and even with extreme reddening assumptions up to 50\%, no statistically significant detection (SNR$>3$) was obtained. In contrast, the August 19 observation reveals a more robust signal: a tentative detection emerges at reddening values above 13\%, and a clear detection (SNR$=3$) is achieved at 24\% reddening.}

\begin{figure*}[ht!]
\centering
\includegraphics[width=\textwidth]{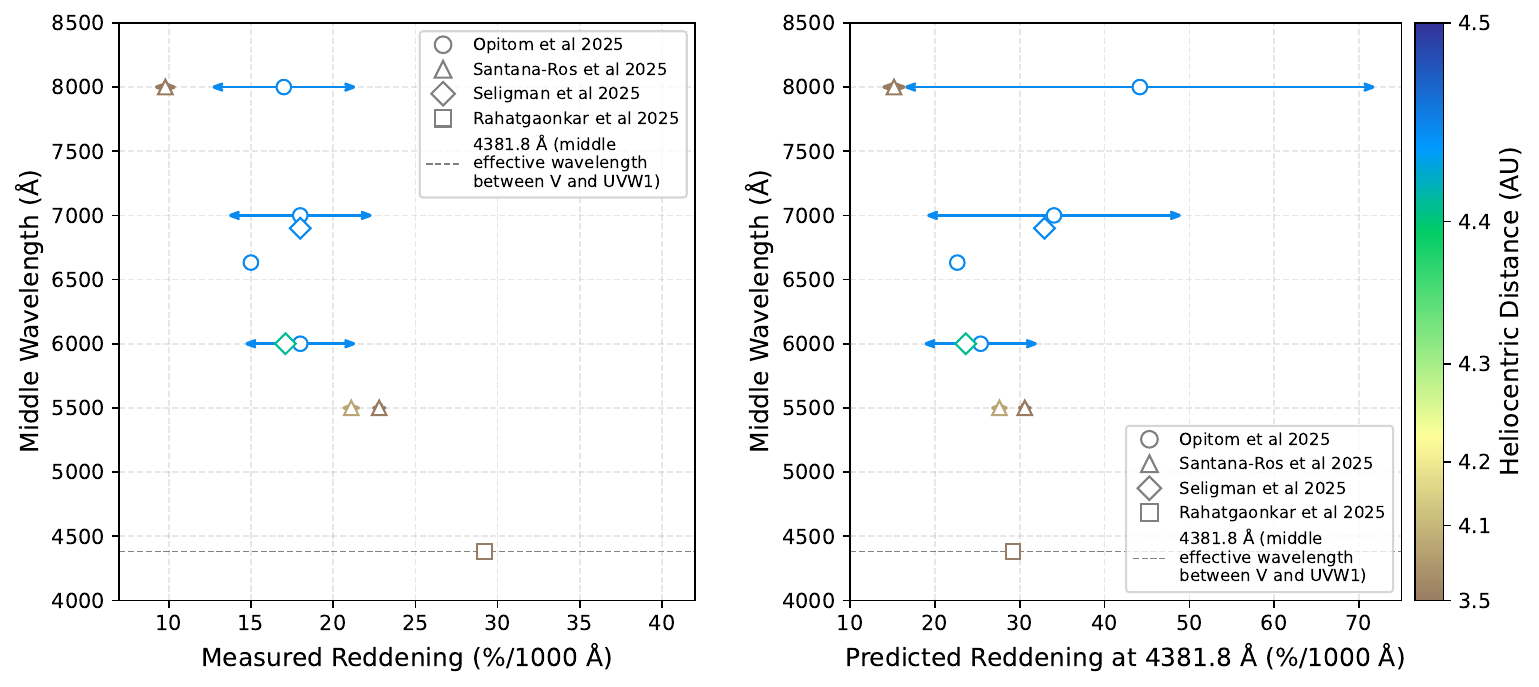}
\caption{Compilation of reddening measurements for 3I/ATLAS. Left panel: Reddening measurements of 3I/ATLAS in literature versus the central wavelength of the measurement bands. Marker shapes denote references; colors indicate heliocentric distance (colorbar at far right). Black dashed line marks the middle effective wavelength between V and UVW1 (4381.8~\AA). Right panel: Same measurements converted to reddening at the middle effective wavelength between V and UVW1. Y-axis shows original measurement wavelengths; x-axis shows converted reddening values at our middle wavelength. Symbols and colorbar as in the left panel.}
\label{fig:reddening}
\end{figure*}

We next examined reddening measurements of 3I/ATLAS obtained with other telescopes, shown in the left panel of Fig.~\ref{fig:reddening}. The x-axis gives reddening values (percent per 1000~\AA), while the y-axis marks the central wavelengths of the respective measurement bands. Marker shapes identify the literature sources, and the colors indicate the comet’s heliocentric distance at the time of observation.

\zx{\citet{rahatgaonkar2025} reported a color slope of approximately $22 \pm 1$\% per 1000~\AA\ between 3900~\AA\ and 5550~\AA, which remained nearly constant from July 4 to August 21. Using their reflectance spectra (obtained via private communication), we measured the reddening (normalized color slope) at 4381.8~\AA\ to be approximately 29\%, which also remained constant through time.}
We show this reddening measurement in Fig.~\ref{fig:qwater} with a black dotted line. Other reddening measurements typically range from 10\% to 25\%, with higher reddening values at shorter wavelengths.

This trend of increasing reddening toward shorter wavelengths is unsurprising. For example, a general trend of increased reddening of cometary dust at shorter central wavelengths is presented in \cite{Jew86}. Directly measured values in the UV band corroborate this trend, with values of 60--80\%/1000~\AA\ near 2950~\AA\ from C/1983 O1 (\v{C}ernis) \citep{FeldmanAhearn1985} and 70\%/1000~\AA\ near 2950~\AA\ from 1P/Halley \citep{Feldman1987}. In contrast, ground-based measurements in the B-V domain hover around 20\% depending on cometary activity, size, and composition \citep{Opitom2025,Storrs1992,Jew86}.

This trend likely arises from two effects. First, reddening is defined as the reflectance spectral slope normalized by the mean reflectance, rather than the raw slope. This normalization introduces wavelength-dependent variations even for linear reflectance spectra, producing higher values at shorter wavelengths. Second, many reflectance spectra are non-linear and typically steepen toward shorter wavelengths, further enhancing reddening values. \zx{Although the reflectance spectrum of 3I/ATLAS does not show such steepening \citep{rahatgaonkar2025}, the first effect alone can account for the trend observed in the left panel.}

To compare literature reddening measures directly with our \emph{Swift} analysis, we converted the reported reddening to the values at effective wavelengths of UVW1 and V (Fig.~\ref{fig:reddening}).  For each measurement, we constructed a linear reflectance spectrum that, when multiplied by the solar spectrum, produces a red dust spectrum with the reported reddening value at that measurement wavelength. We then forward-modeled this reddened spectrum through the transmission curves of \emph{Swift}'s UVW1 and V filters and computed the reddening at UVOT's effective wavelengths. Uncertainties in the converted reddening values were derived using error propagation.

We present the converted results in the right panel of Fig.~\ref{fig:reddening}. The y-axis still shows the original measurement wavelengths, while the x-axis displays the reddening converted to UVOT wavelengths. All reddening values increased after conversion, as expected from the reddening definition. \zx{Nearly all converted values exceed 20\%, with a mean of 29\%}. The conversion accuracy depends on the wavelength difference between the original measurement (y-axis) and UVOT's middle effective wavelength (4381.8~\AA) — smaller differences yield higher accuracy. \zx{The reddening of 29\% measured by \citet{rahatgaonkar2025} at 4381.8~\AA\ is theoretically the most accurate. Additionally, this measurement matches the mean converted value, which not only validates the assumption of a linear reflectance spectrum but also strengthens our confidence in the reddening value.}

\zx{Consequently, we use a reddening of 29\% for subsequent analysis and adopt the corresponding water production rate as our measurement results. For the July 31 visit, the number of OH molecules within the 10-arcsec aperture is $(2.19 \pm 1.47) \times 10^{30}$ (SNR = 1.5), corresponding to a water production rate of $(0.74 \pm 0.50) \times 10^{27}$ molecules s$^{-1}$, indicating a marginal detection. In contrast, the August 19 visit yields $(4.21 \pm 1.07) \times 10^{30}$ OH molecules within the aperture (SNR = 3.9) and a water production rate of $(1.36 \pm 0.35) \times 10^{27}$ molecules s$^{-1}$, and shows a clear detection. A summary of these results is provided in Table~\ref{tab:observation_and_results}.}

\db{Filter imaging of comets, in particular with filters that were not specifically designed for comet observations, can be contaminated by emission features from other gases. Recent observations from JWST and SPHEREx discovered a CO$_2$-dominated coma \citep{cordiner2025, lisse2025}. The ionized product of CO$_2$, CO$_2^+$, has the (B$^2\Sigma_u^+$--X$^2\Pi_g$) doublet between 2884--2896~\AA\ which lies within the bandpass of the UVW1 filter, possibly contaminating our signal. However, sensitive upper limits from the X-shooter spectrum (private communication) show no detectable CO$_2^+$ A–X emission around 3674~\AA. This rules out a significant contribution of the CO$_2^+$ B–X doublet to the UVW1 count rate. Additionally, recent measurements from \citet{rahatgaonkar2025} indicate the presence of Ni emission in the same spectral region. Thus, a potential Ni contribution cannot be ruled out, but quantifying its fraction is highly uncertain with the current data.}

\section{Discussion}

The gas-to-dust ratio, expressed as log$_{10}$[A(0)f$\rho$/Q(H$_2$O)], \zx{was $-24.6$ for both visits}. This value is consistent with solar system comets observed at heliocentric distances of \zx{2.9–3.5 au}, which span a broad range from $-23$ to $-26$ but generally trend toward lower values. The decrease in Af$\rho$-to-water ratios with increasing heliocentric distance is partly attributable to the Af$\rho$ parameter’s assumption of constant outflow \citep{Ahe95}.

Using a sublimation model\footnote{\url{https://pdssbn.astro.umd.edu/tools/ma-evap/index.shtml}} \citep{Cow79} and assuming constant solar elevation (as would occur for slow rotation or a Sun-facing pole), a Bond albedo of 5\%, and unity infrared emissivity, we estimate the minimum required minimal active area to be 10.6~km$^2$ for the first visit and 7.8 km$^2$ for the second visit. Observations acquired with the Hubble Space Telescope placed an upper limit on the nucleus's radius of 2.8~km \citep{Jewitt2025}. If the water is produced directly by sublimation of ice within the nucleus, it would require that 11--8\% of the surface contributes, whereas most solar system comets have active fractions around 3--5\% \citep{Ahe95, Keller2015}.

 \zx{The water production rates measured by \emph{Swift} on July 31 (7.4 $\pm$ 5.0 $\times 10^{26}$ molecules/s) are comparable but slightly higher than the JWST measurement acquired on August 6 (2.19 $\pm$ 0.08 $\times 10^{26}$ molecules/s). This discrepancy may reflect systematic uncertainties if the true H$_2$O and OH distributions deviate from the assumptions of the adopted coma models. Under such circumstances, using different aperture sizes — $\sim$3~\arcsec for JWST and 10~\arcsec for \emph{Swift}—can yield different water production rates. For example, an extended source located outside the JWST aperture but within the \emph{Swift} aperture could contribute additional OH, leading to a higher inferred production rates from the \emph{Swift} measurements.}
\sho{The recent direct detection of water in 3I/ATLAS by JWST similarly shows a growing water production curve with nucleocentric distance, underscoring that nuclear sublimation alone cannot explain observed water activity and distributions at these heliocentric distances.}

 Only a few comets, such as C/1980 E1 (Bowell) and C/2009 P1 (Garradd), have shown OH detections at such large distances outside of 2.5~au. OH emission from Bowell was detected at 5.3~au \citep{AHearn1984}, requiring an implausibly large active area (1.6 $\times$ 10$^7$~km$^2$) unless icy grains contributed significantly to water release. The OH production showed non-monotonic evolution, suggesting transient grain-driven activity. Similarly, \emph{Swift} and SOHO/SWAN observations of comet Garradd revealed elevated water production beyond 3.4~au compared to narrow-slit spectroscopy, consistent with extended sources such as icy grains in the coma \citep{Combi2013, Bod14}.

Near-infrared spectroscopic observations of 3I/ATLAS obtained with Gemini South/GMOS on July 5, 2025 UTC, and IRTF/SpeX on July 14, 2025 UTC, suggest the presence of large icy grains in the coma \citep{Yang2025}. If these grains are the primary source of the water and OH observed in the coma, they may explain why OH (A$^2\Sigma$ -- X$^2\Pi$) emission was detected \zx{nearly simultaneously with} CN emission. Typically, CN is among the most distant fragment species detected in comets due to the high fluorescence efficiency of its B$^2\Sigma^+$ -- X$^2\Sigma^+$ (0,0) violet system near 3883~\AA, which also benefits from lower atmospheric extinction and higher detector sensitivity compared to the OH emission near 3085~\AA\ \citep[c.f.][]{Cochran1991}. Large, dark icy grains in the coma are more efficiently heated than the nucleus, leading to preferential loss of their more volatile components. A similar process was observed at 67P/Churyumov–Gerasimenko, where the transport of material from the southern to the northern hemisphere resulted in enhanced H$_2$O abundances relative to more volatile species such as CO$_2$, CO, and HCN, implying that these more volatile ices were lost as lofted backfall material \citep{Keller2017}.

\jn{With the recent spectral confirmation of CN in the coma of 3I/ATLAS we can derive log$_{10}$(CN/OH) values to compare to typical solar system comet values. Reported CN production rates between August 10 and 19 span a wide range, from 0.2 to 7.8$\times$10$^{24}$ molecules/s \citep{2025ATel17352....1S,rahatgaonkar2025,Manzano2025}. 
To derive $\log_{10}(\mathrm{CN}/\mathrm{OH})$, we obtained the OH production rate ($Q_{\mathrm{OH}}$) by multiplying the measured water production rate by the branching ratio (0.93) and interpolating to each $r_\mathrm{h}$, then computed $\log_{10}(\mathrm{CN}/\mathrm{OH})$ for all CN measurements, yielding a median value of about $-3.1$, which is resilient to outliers.
This would make 3I/ATLAS depleted in CN compared to even depleted solar system comets, which exhibit log$_{10}$(CN/OH)=-2.69 \citep{Ahe95}. 
However, this estimate is highly model-dependent, since $Q_{\mathrm{OH}}$ is derived from water production rates using vectorial model assumptions that may not hold at large heliocentric distances.
}

The potential role of icy grains highlights the complexity of the physical processes governing distant activity and volatile retention or release, which remain poorly constrained due to observational biases, limited statistics, and the difficulty of detecting tenuous outgassing. \zx{Therefore, the assumptions of standard vectorial models — steady-state production, symmetric outflow, and nucleus-centered sources — may break down, leading to systematic uncertainty in the derived water and OH production rates.  In addition, our preliminary radial profiles indicate a more concentrated OH distribution than predicted by the vectorial model, implying that both $Q(\mathrm{H}2\mathrm{O})$ and $Q({\mathrm{OH}})$ may be overestimated. Consequently, the aperture-integrated number of OH molecules provides a more direct and less model-dependent constraint, while the conversion to production rates should be treated with caution.}

The unusual characteristics of 3I/ATLAS including early OH, \jn{a depleted and possibly varying CN/H$_2$O ratio \citep{2025ATel17352....1S,rahatgaonkar2025, Manzano2025}}, strong coma reddening, and inferred large grains, highlight the need for long-term, multiwavelength monitoring. Simultaneous observations of gas and ice will be essential to disentangle the roles of native nucleus sublimation and extended grain activity.

\section{Conclusion}

We report the detection of  OH (A$^2\Sigma$ -- X$^2\Pi$) emission in the coma of 3I/ATLAS using ultraviolet imaging with the \emph{Neil Gehrels-Swift Observatory}. This confirms the presence of water in the third known interstellar object. 
\zx{Based on a dust reddening of 29\% at 4381.8\AA, we obtained only a marginal detection of OH emission during the July 31–August 1 visit at $r_\mathrm{h} = 3.5$~au, corresponding to a water production rate of ($(0.74 \pm 0.50) \times 10^{27}$ molecules\,s$^{-1}$. However,  subsequent observations on August 18–20 revealed a clear detection, corresponding to a water production rate of $(1.36 \pm 0.35) \times 10^{27}$ molecules\,s$^{-1}$ ($\sim$40\,kg\,s$^{-1}$) at $r_\mathrm{h} = 2.9$~au. This is notable given that water sublimation is typically inefficient at such large heliocentric distances.}
This level of activity is consistent with a reasonably-sized active area assuming equilibrium sublimation, but could also result from the presence of large icy grains in the coma, as suggested by contemporaneous near-infrared observations.
If the coma of 3I/ATLAS \jn{becomes} H$_2$O-dominated \jn{during and after perihelion} and \jn{shows lower abundances of higher-metallicity volatiles like CO, CN, and CO$_2$, a trend that would be the inverse of 2I/Borisov, this may support the early and low-metallicity formation hypothesis, and the derived large size of the nucleus could be indicative of a key knowledge gap in low-metallicity system planetesimal formation and loss mechanisms.
} 

The detection of OH emission \jn{nearly simultaneously} to the detection of CN is unusual and may indicate differences in grain composition or volatile distribution compared to solar system comets \citep{Opitom2025}. 
While similar behavior has been observed in comets such as C/1980 E1 (Bowell) and C/2009 P1 (Garradd), the processes governing distant activity and the release of volatiles remain poorly constrained.
Grain-driven activity, in particular, may play a more significant role in interstellar objects than previously assumed. This detection demonstrates the critical role of UV space-based observations in tracking the onset and evolution of activity in distant or dynamically unusual objects.
\jn{The relatively weak CN abundances relative to H$_2$O and the apparent large abundance of CO$_2$ relative to H$_2$O would seem to contradict the theorized low-metallicity formation.}
\jn{The upcoming perihelion passage will likely expose less-altered materials as depleted layers are stripped from its surface, and we can continue to test the low-metallicity hypothesis.} Continued monitoring of 3I/ATLAS will help clarify the thermal and compositional properties of interstellar ices and improve our understanding of volatile retention and release in planetary systems beyond our own.


\begin{acknowledgments}
We thank the \emph{Neil Gehrels-Swift Observatory} team for granting us observing time through the Director’s Discretionary Program. We are especially grateful to Mike Siegel, Sophia Lanava, Maia Williams, and Thomas Gaudin for their careful and effective planning which ensured the success of our observations.
\sho{We also thank the VLT team members Juan Pablo Carvajal, Rohan Rahatgaonkar, Baltasar Luco, and Thomas Puzia for their valuable contributions of time, data, and expertise with spectral observations and dust reddening measurements.}
\sho{Additionally, we thank Steven Bromley for detailed atomic fluorescence modeling support.}
\end{acknowledgments}

\begin{contribution}
All authors contributed equally to the collaboration. ZX was responsible for the design of the observations, conducted the formal analysis, developed the software, and created the visualizations.  
SO contributed to the formal analysis, software development, and visualizations. JN contributed to the conceptual framing and provided contextual expertise.   
DB conceived the proposal, led the overall conceptualization of the study, and managed the project. 
All authors contributed to the writing, review, and editing of the manuscript.
\end{contribution}

%
\facilities{Swift(UVOT)}

\software{astropy~\citep{2013A&A...558A..33A,2018AJ....156..123A,2022ApJ...935..167A},  
sbpy~\citep{Mommert2019},
synphot~\citep{synphot_ASCL_1811},
stsynphot~\citep{stsynphot_ASCL_2010}
          }


\bibliography{main}{}

\begin{thebibliography}{}
\expandafter\ifx\csname natexlab\endcsname\relax\def\natexlab#1{#1}\fi
\providecommand{\url}[1]{\href{#1}{#1}}
\providecommand{\dodoi}[1]{doi:~\href{http://doi.org/#1}{\nolinkurl{#1}}}
\providecommand{\doeprint}[1]{\href{http://ascl.net/#1}{\nolinkurl{http://ascl.net/#1}}}
\providecommand{\doarXiv}[1]{\href{https://arxiv.org/abs/#1}{\nolinkurl{https://arxiv.org/abs/#1}}}

\bibitem[{M.~F. {A'Hearn} {et~al.}(1995){A'Hearn}, {Millis}, {Schleicher}, {Osip}, \& {Birch}}]{Ahe95}
{A'Hearn}, M.~F., {Millis}, R.~C., {Schleicher}, D.~O., {Osip}, D.~J., \& {Birch}, P.~V. 1995, \bibinfo{title}{{The ensemble properties of comets: Results from narrowband photometry of 85 comets, 1976-1992.},} Icarus, 118, 223, \dodoi{10.1006/icar.1995.1190}

\bibitem[{M.~F. A'Hearn {et~al.}(1984)A'Hearn, Schleicher, Millis, Feldman, \& Thompson}]{AHearn1984}
A'Hearn, M.~F., Schleicher, D.~G., Millis, R.~L., Feldman, P.~D., \& Thompson, D.~T. 1984, \bibinfo{title}{{Comet Bowell 1980b},} The Astronomical Journal, 89, 579, \dodoi{10.1086/113552}

\bibitem[{M.~F. A'Hearn {et~al.}(2011)A'Hearn, Belton, Delamere, Feaga, Hampton, Kissel, Klaasen, McFadden, Meech, Melosh, Schultz, Sunshine, Thomas, Veverka, Wellnitz, Yeomans, Besse, Bodewits, Bowling, Carcich, Collins, Farnham, Groussin, Hermalyn, Kelley, Kelley, Li, Lindler, Lisse, McLaughlin, Merlin, Protopapa, Richardson, \& Williams}]{AH2011}
A'Hearn, M.~F., Belton, M. J.~S., Delamere, W.~A., {et~al.} 2011, \bibinfo{title}{{EPOXI at Comet Hartley 2},} Science, 332, 1396, \dodoi{10.1126/science.1204054}

\bibitem[{ {Astropy Collaboration} {et~al.}(2022){Astropy Collaboration}, {Price-Whelan}, {Lim}, \& {Earl}}]{2022ApJ...935..167A}
{Astropy Collaboration}, {Price-Whelan}, A.~M., {Lim}, P.~L., \& {Earl}. 2022, \bibinfo{title}{{The Astropy Project: Sustaining and Growing a Community-oriented Open-source Project and the Latest Major Release (v5.0) of the Core Package},} \apj, 935, 167, \dodoi{10.3847/1538-4357/ac7c74}

\bibitem[{ {Astropy Collaboration} {et~al.}(2013){Astropy Collaboration}, {Robitaille}, {Tollerud}, {Greenfield}, {Droettboom}, {Bray}, {Aldcroft}, {Davis}, {Ginsburg}, {Price-Whelan}, {Kerzendorf}, {Conley}, {Crighton}, {Barbary}, {Muna}, {Ferguson}, {Grollier}, {Parikh}, {Nair}, {Unther}, {Deil}, {Woillez}, {Conseil}, {Kramer}, {Turner}, {Singer}, {Fox}, {Weaver}, {Zabalza}, {Edwards}, {Azalee Bostroem}, {Burke}, {Casey}, {Crawford}, {Dencheva}, {Ely}, {Jenness}, {Labrie}, {Lim}, {Pierfederici}, {Pontzen}, {Ptak}, {Refsdal}, {Servillat}, \& {Streicher}}]{2013A&A...558A..33A}
{Astropy Collaboration}, {Robitaille}, T.~P., {Tollerud}, E.~J., {et~al.} 2013, \bibinfo{title}{{Astropy: A community Python package for astronomy},} \aap, 558, A33, \dodoi{10.1051/0004-6361/201322068}

\bibitem[{ {Astropy Collaboration} {et~al.}(2018){Astropy Collaboration}, {Price-Whelan}, {Sip{\H{o}}cz}, {G{\"u}nther}, {Lim}, {Crawford}, {Conseil}, {Shupe}, {Craig}, {Dencheva}, {Ginsburg}, {VanderPlas}, {Bradley}, {P{\'e}rez-Su{\'a}rez}, {de Val-Borro}, {Aldcroft}, {Cruz}, {Robitaille}, {Tollerud}, {Ardelean}, {Babej}, {Bach}, {Bachetti}, {Bakanov}, {Bamford}, {Barentsen}, {Barmby}, {Baumbach}, {Berry}, {Biscani}, {Boquien}, {Bostroem}, {Bouma}, {Brammer}, {Bray}, {Breytenbach}, {Buddelmeijer}, {Burke}, {Calderone}, {Cano Rodr{\'\i}guez}, {Cara}, {Cardoso}, {Cheedella}, {Copin}, {Corrales}, {Crichton}, {D'Avella}, {Deil}, {Depagne}, {Dietrich}, {Donath}, {Droettboom}, {Earl}, {Erben}, {Fabbro}, {Ferreira}, {Finethy}, {Fox}, {Garrison}, {Gibbons}, {Goldstein}, {Gommers}, {Greco}, {Greenfield}, {Groener}, {Grollier}, {Hagen}, {Hirst}, {Homeier}, {Horton}, {Hosseinzadeh}, {Hu}, {Hunkeler}, {Ivezi{\'c}}, {Jain}, {Jenness}, {Kanarek}, {Kendrew}, {Kern}, {Kerzendorf}, {Khvalko}, {King}, {Kirkby}, {Kulkarni},
  {Kumar}, {Lee}, {Lenz}, {Littlefair}, {Ma}, {Macleod}, {Mastropietro}, {McCully}, {Montagnac}, {Morris}, {Mueller}, {Mumford}, {Muna}, {Murphy}, {Nelson}, {Nguyen}, {Ninan}, {N{\"o}the}, {Ogaz}, {Oh}, {Parejko}, {Parley}, {Pascual}, {Patil}, {Patil}, {Plunkett}, {Prochaska}, {Rastogi}, {Reddy Janga}, {Sabater}, {Sakurikar}, {Seifert}, {Sherbert}, {Sherwood-Taylor}, {Shih}, {Sick}, {Silbiger}, {Singanamalla}, {Singer}, {Sladen}, {Sooley}, {Sornarajah}, {Streicher}, {Teuben}, {Thomas}, {Tremblay}, {Turner}, {Terr{\'o}n}, {van Kerkwijk}, {de la Vega}, {Watkins}, {Weaver}, {Whitmore}, {Woillez}, {Zabalza}, \& {Astropy Contributors}}]{2018AJ....156..123A}
{Astropy Collaboration}, {Price-Whelan}, A.~M., {Sip{\H{o}}cz}, B.~M., {et~al.} 2018, \bibinfo{title}{{The Astropy Project: Building an Open-science Project and Status of the v2.0 Core Package},} \aj, 156, 123, \dodoi{10.3847/1538-3881/aabc4f}

\bibitem[{N. {Biver} {et~al.}(2018){Biver}, {Bockel{\'e}e-Morvan}, {Paubert}, {Moreno}, {Crovisier}, {Boissier}, {Bertrand}, {Boussier}, {Kugel}, {McKay}, {Dello Russo}, \& {DiSanti}}]{Biver2018}
{Biver}, N., {Bockel{\'e}e-Morvan}, D., {Paubert}, G., {et~al.} 2018, \bibinfo{title}{{The extraordinary composition of the blue comet C/2016 R2 (PanSTARRS)},} \aap, 619, A127, \dodoi{10.1051/0004-6361/201833449}

\bibitem[{D. {Bodewits} {et~al.}(2014){Bodewits}, {Farnham}, {A'Hearn}, {Feaga}, {McKay}, {Schleicher}, \& {Sunshine}}]{Bod14}
{Bodewits}, D., {Farnham}, T.~L., {A'Hearn}, M.~F., {et~al.} 2014, \bibinfo{title}{{The Evolving Activity of the Dynamically Young Comet C/2009 P1 (Garradd)},} \apj, 786, 48, \dodoi{10.1088/0004-637X/786/1/48}

\bibitem[{D. Bodewits {et~al.}(2019)Bodewits, Orszagh, Noonan, {\v{D}}urian, \& Matej{\v c}{\'\i}k}]{Bod19}
Bodewits, D., Orszagh, J., Noonan, J., {\v{D}}urian, M., \& Matej{\v c}{\'\i}k, {\v{S}}. 2019, \bibinfo{title}{{Diagnostics of Collisions between Electrons and Water Molecules in Near-ultraviolet and Visible Wavelengths},} Astrophysical Journal, 885, 167

\bibitem[{D. Bodewits {et~al.}(2023)Bodewits, Xing, Saki, \& Morgenthaler}]{bodewits2023}
Bodewits, D., Xing, Z., Saki, M., \& Morgenthaler, J.~P. 2023, \bibinfo{title}{{Neil Gehrels-Swift Observatory's Ultraviolet/Optical Telescope Observations of Small Bodies in the Solar System},} Universe, 78, \dodoi{10.3390/universe9020078}

\bibitem[{D. Bodewits {et~al.}(2020)Bodewits, Noonan, Feldman, Bannister, Farnocchia, Harris, Li, Mandt, Parker, \& Xing}]{Bodewits2020}
Bodewits, D., Noonan, J.~W., Feldman, P.~D., {et~al.} 2020, \bibinfo{title}{{The carbon monoxide-rich interstellar comet 2I/Borisov},} Nature Astronomy, 4, 867, \dodoi{10.1038/s41550-020-1095-2}

\bibitem[{R.~C. Bohlin(2014)Bohlin}]{Bohlin2014_CALSPEC}
Bohlin, R.~C. 2014, \bibinfo{title}{Hubble Space Telescope CALSPEC Flux Standards: Sirius (and Vega),} AJ, 147, 127, \dodoi{10.1088/0004-6256/147/6/127}

\bibitem[{A.~A. Breeveld {et~al.}(2010)Breeveld, Curran, Hoversten, Koch, Landsman, Marshall, Page, Poole, Roming, Smith, Still, Yershov, Blustin, Brown, Gronwall, Holland, Kuin, McGowan, Rosen, Boyd, Broos, Carter, Chester, Hancock, Huckle, Immler, Ivanushkina, Kennedy, Mason, Morgan, Oates, De~Pasquale, Schady, Siegel, \& Vanden~Berk}]{Breeveld_2010}
Breeveld, A.~A., Curran, P.~A., Hoversten, E.~A., {et~al.} 2010, \bibinfo{title}{Further calibration of the Swift ultraviolet/optical telescope: Further calibration of the Swift UVOT,} Monthly Notices of the Royal Astronomical Society, \dodoi{10.1111/j.1365-2966.2010.16832.x}

\bibitem[{A.~A. Breeveld {et~al.}(2011)Breeveld, Landsman, Holland, Roming, Kuin, Page, McEnery, Racusin, \& Gehrels}]{Breeveld_2011_sensitivity}
Breeveld, A.~A., Landsman, W., Holland, S.~T., {et~al.} 2011, \bibinfo{title}{An Updated Ultraviolet Calibration for the Swift/UVOT,} in An Updated Ultraviolet Calibration for the Swift/UVOT, Annapolis, MD, (USA), 373–376, \dodoi{10.1063/1.3621807}

\bibitem[{J.~A. Carter {et~al.}(2012)Carter, Bodewits, Read, \& Immler}]{Carter2012}
Carter, J.~A., Bodewits, D., Read, A.~M., \& Immler, S.~M. 2012, \bibinfo{title}{{Simultaneous Swift X-ray and UV views of comet C/2007 N3 (Lulin)},} Astronomy and Astrophysics, 541, 70, \dodoi{10.1051/0004-6361/201117950}

\bibitem[{A.~L. Cochran \& W.~D. Cochran(1991)Cochran \& Cochran}]{Cochran1991}
Cochran, A.~L., \& Cochran, W.~D. 1991, \bibinfo{title}{{The first detection of CN and the distribution of CO(+) gas in the coma of Comet P/Schwassmann-Wachmann 1},} Icarus, 90, 172, \dodoi{10.1016/0019-1035(91)90077-7}

\bibitem[{M.~R. Combi {et~al.}(2004)Combi, Harris, \& Smyth}]{Com04}
Combi, M.~R., Harris, W.~M., \& Smyth, W.~H. 2004, \bibinfo{title}{{Gas dynamics and kinetics in the cometary coma: theory and observations},} in Comets II, ed. H.~A. Weaver, H.~U. Keller, \& M.~Festou (Tuscon, AZ: University of Arizona Press), 523

\bibitem[{M.~R. Combi {et~al.}(2013)Combi, Makinen, Bertaux, Quemerais, Ferron, \& Fougere}]{Combi2013}
Combi, M.~R., Makinen, J. T.~T., Bertaux, J.-L., {et~al.} 2013, \bibinfo{title}{{Water production rate of Comet C/2009 P1 (Garradd) throughout the 2011-2012 apparition: Evidence for an icy grain halo},} Icarus, 225, 740 , \dodoi{10.1016/j.icarus.2013.04.030}

\bibitem[{M.~A. Cordiner {et~al.}(2020)Cordiner, Milam, Biver, e~Morvan, Roth, Bergin, Jehin, Remijan, Charnley, Mumma, Boissier, Crovisier, Paganini, Kuan, \& Lis}]{Cordiner2020}
Cordiner, M.~A., Milam, S.~N., Biver, N., {et~al.} 2020, \bibinfo{title}{{Unusually high CO abundance of the first active interstellar comet},} Nature Astronomy, 1 , \dodoi{10.1038/s41550-020-1087-2}

\bibitem[{M.~A. {Cordiner} {et~al.}(2025){Cordiner}, {Roth}, {Kelley}, {Bodewits}, {Charnley}, {Drozdovskaya}, {Farnocchia}, {Micheli}, {Milam}, {Opitom}, {Schwamb}, \& {Thomas}}]{cordiner2025}
{Cordiner}, M.~A., {Roth}, N.~X., {Kelley}, M. S.~P., {et~al.} 2025, \bibinfo{title}{{JWST detection of a carbon dioxide dominated gas coma surrounding interstellar object 3I/ATLAS},} arXiv e-prints, arXiv:2508.18209, \dodoi{10.48550/arXiv.2508.18209}

\bibitem[{J.~J. {Cowan} \& M.~F. {A'Hearn}(1979){Cowan} \& {A'Hearn}}]{Cow79}
{Cowan}, J.~J., \& {A'Hearn}, M.~F. 1979, \bibinfo{title}{{Vaporization of comet nuclei: Light curves and life times},} Moon and Planets, 21, 155, \dodoi{10.1007/BF00897085}

\bibitem[{L. {Denneau}(2025){Denneau}}]{MPC2025K25N12}
{Denneau}, L. 2025, {Minor Planet Electronic Circular K25N12},, \url{https://minorplanetcenter.net/mpec/K25/K25N12.html}

\bibitem[{P.~D. Feldman \& M.~F. A'Hearn(1985)Feldman \& A'Hearn}]{FeldmanAhearn1985}
Feldman, P.~D., \& A'Hearn, M.~F. 1985, \bibinfo{title}{Ultraviolet Albedo of Cometary Grains,} in NATO ASI Series C, Vol. 156, Ices in the Solar System, ed. J.~Klinger, D.~Benest, A.~Dollfus, \& R.~Smoluchowski (Dordrecht: D. Reidel (Springer)), 453--461, \dodoi{10.1007/978-94-009-5418-2_31}

\bibitem[{P.~D. {Feldman} {et~al.}(1987){Feldman}, {Festou}, {A'Hearn}, {Arpigny}, {Butterworth}, {Cosmovici}, {Danks}, {Gilmozzi}, {Jackson}, {McFadden}, {Patriarchi}, {Schleicher}, {Tozzi}, {Wallis}, {Weaver}, \& {Woods}}]{Feldman1987}
{Feldman}, P.~D., {Festou}, M.~C., {A'Hearn}, M.~F., {et~al.} 1987, \bibinfo{title}{{IUE Observations of Comet p/ Halley - Evolution of the Ultraviolet Spectrum Between 1985SEP and 1986JUL},} \aap, 187, 325

\bibitem[{M.~C. Festou(1981)Festou}]{Fes81}
Festou, M.~C. 1981, \bibinfo{title}{{The density distribution of neutral compounds in cometary atmospheres. I - Models and equations},} Astronomy and Astrophysics, 95, 69

\bibitem[{A. {Fitzsimmons} {et~al.}(2023){Fitzsimmons}, {Meech}, {Matr{\`a}}, \& {Pfalzner}}]{Fitzsimmons2023}
{Fitzsimmons}, A., {Meech}, K., {Matr{\`a}}, L., \& {Pfalzner}, S. 2023, \bibinfo{title}{{Interstellar Objects and Exocomets},} arXiv e-prints, arXiv:2303.17980, \dodoi{10.48550/arXiv.2303.17980}

\bibitem[{N. {Gehrels} {et~al.}(2004){Gehrels}, {Chincarini}, {Giommi}, {Mason}, {Nousek}, {Wells}, {White}, {Barthelmy}, {Burrows}, {Cominsky}, {Hurley}, {Marshall}, {M{\'e}sz{\'a}ros}, {Roming}, {Angelini}, {Barbier}, {Belloni}, {Campana}, {Caraveo}, {Chester}, {Citterio}, {Cline}, {Cropper}, {Cummings}, {Dean}, {Feigelson}, {Fenimore}, {Frail}, {Fruchter}, {Garmire}, {Gendreau}, {Ghisellini}, {Greiner}, {Hill}, {Hunsberger}, {Krimm}, {Kulkarni}, {Kumar}, {Lebrun}, {Lloyd-Ronning}, {Markwardt}, {Mattson}, {Mushotzky}, {Norris}, {Osborne}, {Paczynski}, {Palmer}, {Park}, {Parsons}, {Paul}, {Rees}, {Reynolds}, {Rhoads}, {Sasseen}, {Schaefer}, {Short}, {Smale}, {Smith}, {Stella}, {Tagliaferri}, {Takahashi}, {Tashiro}, {Townsley}, {Tueller}, {Turner}, {Vietri}, {Voges}, {Ward}, {Willingale}, {Zerbi}, \& {Zhang}}]{Gehrels2004}
{Gehrels}, N., {Chincarini}, G., {Giommi}, P., {et~al.} 2004, \bibinfo{title}{{The Swift Gamma-Ray Burst Mission},} \apj, 611, 1005, \dodoi{10.1086/422091}

\bibitem[{M.~J. {Hopkins} {et~al.}(2025){Hopkins}, {Dorsey}, {Forbes}, {Bannister}, {Lintott}, \& {Leicester}}]{hopkins2025different}
{Hopkins}, M.~J., {Dorsey}, R.~C., {Forbes}, J.~C., {et~al.} 2025, \bibinfo{title}{{From a Different Star: 3I/ATLAS in the context of the {\={O}}tautahi-Oxford interstellar object population model},} arXiv e-prints, arXiv:2507.05318, \dodoi{10.48550/arXiv.2507.05318}

\bibitem[{D. {Jewitt} {et~al.}(2025){Jewitt}, {Hui}, {Mutchler}, {Kim}, \& {Agarwal}}]{Jewitt2025}
{Jewitt}, D., {Hui}, M.-T., {Mutchler}, M., {Kim}, Y., \& {Agarwal}, J. 2025, \bibinfo{title}{{Hubble Space Telescope Observations of the Interstellar Interloper 3I/ATLAS},} arXiv e-prints, arXiv:2508.02934.
\newblock \doarXiv{2508.02934}

\bibitem[{D. Jewitt \& K.~J. Meech(1986)Jewitt \& Meech}]{Jew86}
Jewitt, D., \& Meech, K.~J. 1986, \bibinfo{title}{Cometary grain scattering versus wavelength, or “What color is comet dust”?} The Astrophysical Journal, 310, 937, \dodoi{10.1086/164745}

\bibitem[{H.~U. Keller {et~al.}(2015)Keller, Mottola, Davidsson, Schröder, Skorov, Kührt, Groussin, Pajola, Hviid, Preusker, Scholten, A'Hearn, Sierks, Barbieri, Lamy, Rodrigo, Koschny, Rickman, Barucci, Bertaux, Bertini, Cremonese, Deppo, Debei, Cecco, Fornasier, Fulle, Gutierrez, Ip, Jorda, Knollenberg, Kramm, Kuppers, Lara, Lazzarin, Moreno, Marzari, Michalik, Naletto, Sabau, Thomas, Vincent, Wenzel, Agarwal, Guttler, Oklay, \& Tubiana}]{Keller2015}
Keller, H.~U., Mottola, S., Davidsson, B. J.~R., {et~al.} 2015, \bibinfo{title}{{Insolation, erosion, and morphology of comet 67P/Churyumov-Gerasimenko},} Astronomy and Astrophysics, 583, A34, \dodoi{10.1051/0004-6361/201525964}

\bibitem[{H.~U. Keller {et~al.}(2017)Keller, Mottola, Hviid, Agarwal, Kührt, Skorov, Otto, Vincent, Oklay, Schröder, Davidsson, Pajola, Shi, Bodewits, Toth, Preusker, Scholten, Sierks, Barbieri, Lamy, Rodrigo, Koschny, Rickman, A'Hearn, Barucci, Bertaux, Bertini, Cremonese, Deppo, Debei, Cecco, Deller, Fornasier, Fulle, Groussin, Gutierrez, Guttler, Hofmann, Ip, Jorda, Knollenberg, Kramm, Kuppers, Lara, Lazzarin, Moreno, Marzari, Naletto, Tubiana, \& Thomas}]{Keller2017}
Keller, H.~U., Mottola, S., Hviid, S.~F., {et~al.} 2017, \bibinfo{title}{{Seasonal Mass Transfer on the Nucleus of Comet 67P/Chuyumov-Gerasimenko},} Monthly Notices of the Royal Astronomical Society: Letters, 469, S357 , \dodoi{10.1093/mnras/stx1726}

\bibitem[{C.~M. {Lisse} {et~al.}(2025){Lisse}, {Bach}, {Bryan}, {Crill}, {Cukierman}, {Dor{\'e}}, {Fabinsky}, {Faisst}, {Korngut}, {Melnick}, {Rustamkulov}, {Tolls}, {Werner}, {Sitko}, {Champagne}, {Connelley}, {Emery}, {Fernandez}, {Yang}, \& {the SPHEREx Science Team}}]{lisse2025}
{Lisse}, C.~M., {Bach}, Y.~P., {Bryan}, S., {et~al.} 2025, \bibinfo{title}{{SPHEREx Discovery of Strong Water Ice Absorption and an Extended Carbon Dioxide Coma in 3I/ATLAS},} arXiv e-prints, arXiv:2508.15469, \dodoi{10.48550/arXiv.2508.15469}

\bibitem[{K. Mason {et~al.}(2007)Mason, Chester, Cucchiara, Gronwall, Grupe, Hunsberger, Jones, Koch, Nousek, O'Brien, Racusin, Roming, Smith, Wells, Willingale, Branduardi-Raymont, \& Gehrels}]{Mason2007}
Mason, K., Chester, M., Cucchiara, A., {et~al.} 2007, \bibinfo{title}{{Swift ultraviolet photometry of the Deep Impact encounter with Comet 9P/Tempel 1},} Icarus, 187, 123, \dodoi{10.1016/j.icarus.2006.09.021}

\bibitem[{A.~J. {McKay} {et~al.}(2019){McKay}, {DiSanti}, {Kelley}, {Knight}, {Womack}, {Wierzchos}, {Harrington Pinto}, {Bonev}, {Villanueva}, {Dello Russo}, {Cochran}, {Biver}, {Bauer}, {Vervack}, {Gibb}, {Roth}, \& {Kawakita}}]{McKay2019}
{McKay}, A.~J., {DiSanti}, M.~A., {Kelley}, M. S.~P., {et~al.} 2019, \bibinfo{title}{{The Peculiar Volatile Composition of CO-dominated Comet C/2016 R2 (PanSTARRS)},} \aj, 158, 128, \dodoi{10.3847/1538-3881/ab32e4}

\bibitem[{M. Mommert {et~al.}(2019)Mommert, Kelley, Val-Borro, Li, Guzman, Sipőcz, Ďurech, Granvik, Grundy, Moskovitz, Penttilä, \& Samarasinha}]{Mommert2019}
Mommert, M., Kelley, M. S.~p., Val-Borro, M.~d., {et~al.} 2019, \bibinfo{title}{sbpy: A Python module for small-body planetary astronomy,} Journal of Open Source Software, 4, 1426, \dodoi{10.21105/joss.01426}

\bibitem[{T. Ootsubo {et~al.}(2012)Ootsubo, Kawakita, Hamada, Kobayashi, Yamaguchi, Usui, Nakagawa, Ueno, Ishiguro, Sekiguchi, Watanabe, Sakon, Shimonishi, \& Onaka}]{Ootsubo2012}
Ootsubo, T., Kawakita, H., Hamada, S., {et~al.} 2012, \bibinfo{title}{{AKARI Near-infrared Spectroscopic Survey for CO2 in 18 Comets},} Astrophysical Journal, 752, 15, \dodoi{10.1088/0004-637x/752/1/15}

\bibitem[{C. {Opitom} {et~al.}(2025){Opitom}, {Snodgrass}, {Jehin}, {Bannister}, {Bufanda}, {Deam}, {Dorsey}, {Ferrais}, {Hmiddouch}, {Knight}, {Kokotanekova}, {Leicester}, {Marsset}, {Murphy}, {Okoth}, {Ridden-Harper}, {Vander Donckt}, {Ferellec}, {Hutsemekers}, {Lippi}, {Manfroid}, \& {Benkhaldoun}}]{Opitom2025}
{Opitom}, C., {Snodgrass}, C., {Jehin}, E., {et~al.} 2025, \bibinfo{title}{{Snapshot of a new interstellar comet: 3I/ATLAS has a red and featureless spectrum},} arXiv e-prints, arXiv:2507.05226, \dodoi{10.48550/arXiv.2507.05226}

\bibitem[{ {`Oumuamua Team}(2019){`Oumuamua Team}}]{Team2019}
{`Oumuamua Team}. 2019, \bibinfo{title}{{The Natural History of 'Oumuamua},} arXiv.org, astro-ph.EP, 594 , \dodoi{10.1038/s41550-019-0816-x}

\bibitem[{O.~H. Pinto {et~al.}(2022)Pinto, Womack, Fernandez, \& Bauer}]{Pinto2022}
Pinto, O.~H., Womack, M., Fernandez, Y.~R., \& Bauer, J. 2022, \bibinfo{title}{{A Survey of CO, CO2, and H2O in Comets and Centaurs},} arXiv, \dodoi{10.48550/arxiv.2209.09985}

\bibitem[{T.~S. Poole {et~al.}(2008)Poole, Breeveld, Page, Landsman, Holland, Roming, Kuin, Brown, Gronwall, Hunsberger, Koch, Mason, Schady, Berk, Blustin, Boyd, Broos, Carter, Chester, Cucchiara, Hancock, Huckle, Immler, Ivanushkina, Kennedy, Marshall, Morgan, Pandey, Pasquale, Smith, \& Still}]{Poole2008}
Poole, T.~S., Breeveld, A.~A., Page, M.~J., {et~al.} 2008, \bibinfo{title}{{Photometric calibration of the Swift ultraviolet/optical telescope},} Monthly Notices of the Royal Astronomical Society, 383, 627 , \dodoi{10.1111/j.1365-2966.2007.12563.x}

\bibitem[{R. {Rahatgaonkar} {et~al.}(2025){Rahatgaonkar}, {Carvajal}, {Puzia}, {Luco}, {Jehin}, {Hutsem{\'e}kers}, {Opitom}, {Manfroid}, {Marsset}, {Yang}, {Buchanan}, {Fraser}, {Forbes}, {Bannister}, {Bodewits}, {Bolin}, {Belyakov}, {Knight}, {Snodgrass}, {Bufanda}, {Dorsey}, {Ferellec}, {La Forgia}, {Lippi}, {Murphy}, {Nayak}, \& {Vander Donckt}}]{rahatgaonkar2025}
{Rahatgaonkar}, R., {Carvajal}, J.~P., {Puzia}, T.~H., {et~al.} 2025, \bibinfo{title}{{VLT observations of interstellar comet 3I/ATLAS II. From quiescence to glow: Dramatic rise of Ni I emission and incipient CN outgassing at large heliocentric distances},} arXiv e-prints, arXiv:2508.18382.
\newblock \doarXiv{2508.18382}

\bibitem[{P.~W. {Roming} {et~al.}(2000){Roming}, {Townsley}, {Nousek}, {Altimore}, {Case}, {Hunsberger}, {Koch}, {Mason}, {Carter}, {Cropper}, {Hancock}, {Huckle}, {Kennedy}, {McLelland}, {Smith}, {Killough}, \& {Ho}}]{Roming2000}
{Roming}, P.~W., {Townsley}, L.~K., {Nousek}, J.~A., {et~al.} 2000, \bibinfo{title}{{Ultraviolet/Optical Telescope of the Swift MIDEX mission},} in Society of Photo-Optical Instrumentation Engineers (SPIE) Conference Series, Vol. 4140, X-Ray and Gamma-Ray Instrumentation for Astronomy XI, ed. K.~A. {Flanagan} \& O.~H. {Siegmund}, 76--86, \dodoi{10.1117/12.409161}

\bibitem[{L.~E. {Salazar Manzano} {et~al.}(2025){Salazar Manzano}, {Lin}, {Taylor}, {Seligman}, {Adams}, {Gerdes}, {Ruch}, {Frincke}, \& {Napier}}]{Manzano2025}
{Salazar Manzano}, L.~E., {Lin}, H.~W., {Taylor}, A.~G., {et~al.} 2025, \bibinfo{title}{{Onset of CN Emission in 3I/ATLAS: Evidence for Strong Carbon-Chain Depletion},} arXiv e-prints, arXiv:2509.01647.
\newblock \doarXiv{2509.01647}

\bibitem[{T. Santana-Ros {et~al.}(2025)Santana-Ros, Ivanova, Mykhailova, Erasmus, Kami{\'k}ski, Oszkiewicz, Kwiatkowski, Husárik, Ngwane, \& Penttilä}]{SantanaRos2025}
Santana-Ros, T., Ivanova, O., Mykhailova, S., {et~al.} 2025, \bibinfo{title}{Temporal Evolution of the Third Interstellar Comet 3I/ATLAS: Spin, Color, Spectra and Dust Activity,} arXiv, \dodoi{10.48550/arXiv.2508.00808}

\bibitem[{D. {Schleicher}(2025){Schleicher}}]{2025ATel17352....1S}
{Schleicher}, D. 2025, \bibinfo{title}{{The Detection of CN in Interstellar Comet 3I/ATLAS},} The Astronomer's Telegram, 17352, 1

\bibitem[{D.~G. Schleicher \& M.~F. A'Hearn(1988)Schleicher \& A'Hearn}]{Sch88}
Schleicher, D.~G., \& A'Hearn, M.~F. 1988, \bibinfo{title}{{The fluorescence of cometary OH},} Astrophysical Journal, 331, 1058

\bibitem[{D.~G. {Schleicher} \& A.~N. {Bair}(2011){Schleicher} \& {Bair}}]{Schleicher2011}
{Schleicher}, D.~G., \& {Bair}, A.~N. 2011, \bibinfo{title}{{The Composition of the Interior of Comet 73P/Schwassmann-Wachmann 3: Results from Narrowband Photometry of Multiple Components},} \aj, 141, 177, \dodoi{10.1088/0004-6256/141/6/177}

\bibitem[{D.~Z. {Seligman} {et~al.}(2025){Seligman}, {Micheli}, {Farnocchia}, {Denneau}, {Noonan}, {Hsieh}, {Santana-Ros}, {Tonry}, {Auchettl}, {Conversi}, {Devog{\`e}le}, {Faggioli}, {Feinstein}, {Fenucci}, {Ferrais}, {Frincke}, {Gillon}, {Hainaut}, {Hart}, {Hoffman}, {Holt}, {Hoogendam}, {Huber}, {Jehin}, {Kareta}, {Keane}, {Kelley}, {Lister}, {Mandt}, {Manfroid}, {Mar{\v{c}}eta}, {Meech}, {Amine Miftah}, {Morgan}, {Oca{\~n}a}, {Pe{\~n}a-Asensio}, {Shappee}, {Siverd}, {Taylor}, {Tucker}, {Wainscoat}, {Weryk}, {Wray}, {Yaginuma}, {Yang}, {Ye}, \& {Zhang}}]{Seligman2025}
{Seligman}, D.~Z., {Micheli}, M., {Farnocchia}, D., {et~al.} 2025, \bibinfo{title}{{Discovery and Preliminary Characterization of a Third Interstellar Object: 3I/ATLAS},} arXiv e-prints, arXiv:2507.02757, \dodoi{10.48550/arXiv.2507.02757}

\bibitem[{A.~D. {Storrs} {et~al.}(1992){Storrs}, {Cochran}, \& {Barker}}]{Storrs1992}
{Storrs}, A.~D., {Cochran}, A.~L., \& {Barker}, E.~S. 1992, \bibinfo{title}{{Spectrophotometry of the continuum in 18 comets},} \icarus, 98, 163, \dodoi{10.1016/0019-1035(92)90087-N}

\bibitem[{ {STScI Development Team}(2018){STScI Development Team}}]{synphot_ASCL_1811}
{STScI Development Team}. 2018, synphot: Synthetic photometry using Astropy,, Astrophysics Source Code Library, record ascl:1811.001 \url{https://ascl.net/1811.001}

\bibitem[{ {STScI Development Team}(2020){STScI Development Team}}]{stsynphot_ASCL_2010}
{STScI Development Team}. 2020, stsynphot: synphot for HST and JWST,, Astrophysics Source Code Library, record ascl:2010.003 \url{https://ascl.net/2010.003}

\bibitem[{A.~G. {Taylor} \& D.~Z. {Seligman}(2025){Taylor} \& {Seligman}}]{taylor2025kinematic}
{Taylor}, A.~G., \& {Seligman}, D.~Z. 2025, \bibinfo{title}{{The Kinematic Age of 3I/ATLAS and its Implications for Early Planet Formation},} arXiv e-prints, arXiv:2507.08111, \dodoi{10.48550/arXiv.2507.08111}

\bibitem[{Z. Xing {et~al.}(2020)Xing, Bodewits, Noonan, \& Bannister}]{Xing2020}
Xing, Z., Bodewits, D., Noonan, J., \& Bannister, M.~T. 2020, \bibinfo{title}{{Water Production Rates and Activity of Interstellar Comet 2I/Borisov},} The Astrophysical Journal Letters, 893, L48, \dodoi{10.3847/2041-8213/ab86be}

\bibitem[{B. {Yang} {et~al.}(2025){Yang}, {Meech}, {Connelley}, \& {Keane}}]{Yang2025}
{Yang}, B., {Meech}, K.~J., {Connelley}, M., \& {Keane}, J.~V. 2025, \bibinfo{title}{{Spectroscopic Characterization of Interstellar Object 3I/ATLAS: Water Ice in the Coma},} arXiv e-prints, arXiv:2507.14916, \dodoi{10.48550/arXiv.2507.14916}

\end{thebibliography}
\bibliographystyle{aasjournalv7}



\end{document}